\documentclass[secnumarabic,twocolumn,aps,prl,showpacs,showkey,superscriptaddress,prl,floatfix]{revtex4-1}

\usepackage{amsmath,amssymb,amsthm}

\usepackage{hyperref}
\usepackage{amsthm}
\usepackage{graphicx}
\usepackage{subfigure}
\usepackage{color}
\usepackage[centerlast,belowskip=-12pt,aboveskip=2pt]{caption}

\newcommand{\reviewThree}[1]{{{#1}}}

\newcommand{\R}{\mathbb{R}}

\newcommand{\erf}{\mathrm{erf}}

\newcommand{\ta}{\tau_\alpha}

\theoremstyle{plain}

\theoremstyle{definition}

\begin{document}

\title{Heterogeneous connections induce oscillations in large scale networks}

\author{Geoffroy Hermann}%
\affiliation{NeuroMathComp Laboratory, INRIA/ENS Paris}
\author{Jonathan Touboul}
\email[]{ jonathan.touboul@college-de-france.fr}
\affiliation{The Mathematical Neuroscience Laboratory, CIRB / Coll\`ege de France (CNRS UMR 7241, INSERM U1050, UPMC ED 158, MEMOLIFE PSL*)}
\affiliation{BANG Laboratory, INRIA Paris}

\date{\today}%
\begin{abstract}
	Realistic large-scale networks display an heterogeneous distribution of connectivity weights, that might also randomly vary in time. We show that depending on the level of heterogeneity in the connectivity coefficients, different qualitative macroscopic and microscopic regimes emerge. We evidence in particular generic transitions from stationary to perfectly periodic phase-locked regimes as the disorder parameter is increased, both in a simple model treated analytically and in a biologically relevant network made of excitable cells. 
\end{abstract}

\pacs{
89.75.-k, 
87.18.Sn, 
87.19.lm, 
05.70.Ln. 
}
\keywords{Randomly connected neural networks, noise-induced transitions, mean-field limits}
\maketitle
Realistic networks modeling physical or biological systems generally display highly heterogeneous connections~\cite{barrat}. 
A paradigmatic example is given by the distribution of synaptic weights in the central nervous system~\cite{parker:03,marder-goaillard:06}: \\
 \indent (i) the precise number of receptors and the extremely slow plasticity mechanisms induce a static disorder termed \emph{quenched synaptic heterogeneities}. \\
 \indent (ii) Thermal noise, channel noise and the intrinsically noisy mechanisms of release and binding of neurotransmitter~\cite{aldo-faisal:08} result in stochastic variations of the synaptic weights termed \emph{stochastic synaptic noise}. \\
Experimental studies of cortical areas~\cite{aradi-soltesz:02} showed that the degree of heterogeneity in the connections significantly impacts the input-output function, rhythmicity and synchrony. Yet, qualitative effects induced by connection heterogeneities are still poorly understood theoretically. One notable exception is the work of Sompolinsky and collaborators~\cite{sompolinsky-crisanti-etal:88}. In the thermodynamic limit of a one-population firing-rate neuronal network with synaptic weights modeled as centered independent Gaussian random variables, they evidenced a phase transition between a stationary and a chaotic regime as the disorder is increased. 
Besides stationary and chaotic regimes, synchronized oscillations is a highly relevant macroscopic state. It is related to fundamental cortical functions such as memory, attention, sleep and consciousness, and its impairments relate to serious pathologies such as epilepsy or Parkinson's disease~\cite{uhlhaas-singer:06,buszaki:06}.

In this Letter, we show that both quenched and stochastic heterogeneities induce the emergence of macroscopic, perfectly periodic oscillations in the mean-field limit, corresponding to phase-locked oscillations of all neurons in the network. A generic transition from a stationary to a periodic regime as disorder increases will first be demonstrated in the case of firing-rate neurons. Similar reasoning will lead us to uncover the same transition in a more realistic network featuring excitable cells. 

Models of cortical areas involve at least two neural populations, one excitatory corresponding e.g. to pyramidal neurons and one inhibitory population modeling interneurons. In the firing-rate model with $N$ neurons and $P$ populations, the dynamics of the membrane potential $(V^i, i\in\{1\cdots N\})$ of all neurons in the network is given by the system equations:
\vspace{-0.3cm}
\begin{equation*}
\dot{V}^{i}_t=-\frac{V^{i}_t}{\tau_{p_i}}+\sum_{j=1}^{N} J_{ij} S(V^{j}_t)+I_{p_i}(t), \quad i\in\{1\cdots N\}
\end{equation*}

\vspace{-0.3cm}

\noindent where $p_i$ is the population index of neuron $i$, $\tau_{p_i}$ is a common characteristic time of all neurons of population $p_i$ and $I_{p_i}(t)$ their deterministic input. The interaction with the other neurons in the network is given by the sum of a synaptic efficacy $J_{ij}$ multiplied by the firing rate of the presynaptic cell, a positive sigmoidal transform ($S(\cdot)$) of the voltage. Network heterogeneities manifest through random variations of the synaptic coefficients $J_{ij}$ between neuron $i$ and $j$. The two types of heterogeneities in the connections are modeled as follows:\\

\vspace{-0.4cm}
(i) \emph{quenched synaptic heterogeneities}: the $J_{ij}$ are independent {\bf random variables} with law
$\mathcal{N}(\bar{J}_{p_ip_j}/N_{p_j},\sigma/\sqrt{N_{p_j}})$, where $N_{\alpha}$ denotes the number of neurons in population $\alpha\in\{1\cdots P\}$.\\

\vspace{-0.4cm}
(ii) \emph{stochastic synaptic noise}: $J_{ij}$ are {\bf stochastic processes}, e.g. only depending on $i$ and $p_j$ and given by
$J_{ij}=\bar{J}_{p_ip_j}/N_{p_j}+\sigma/N_{p_j} \, \dot{W}^{i p_j}_t$ where the collection $(\dot{W}^{k \alpha}_t, k=1\cdots N, \alpha=1\cdots P)$ are independent Gaussian white noise.

In both case, the scaling ensures that the disorder in the system is quantified by the parameter $\sigma$ and the effective input by the coefficients $(\bar{J}_{\alpha\beta}, \alpha,\beta \in \{1\cdots P\})$. These two models lead to relatively different network equations: model (i) corresponds to a system of $N$ coupled ODEs with random coefficients, whereas model (ii) corresponds to SDEs. Incidentally, the analysis of model (ii) will be analytically simpler and give great insight on the dynamics of system (i).

In the stochastic synaptic noise case, the network satisfies the system of Langevin It\^o equation:
\vspace{-0.2cm}
\begin{equation*}
\dot{V}^{i}_t = -\frac {V^{i}_t} {\tau_{p_i}}   + I_{p_i}(t) + \sum_{\beta=1}^{P} \frac{1}{N_{\beta}} \left(\bar{J}_{p_i\beta}+ \sigma \dot{W}^{i\beta}_{t}\right) \sum_{\{j: p_j=\beta\}} S(V^{j}_t) 
\end{equation*}
These equations are evocative of the kinetic theory of gases, and the mean-field theory~\cite{sznitman:89} allows rigorously demonstrating, in the limit where all the $N_{\beta}$ tend to infinity, the propagation of chaos property and the convergence in law of $V^i_t$ for $p_i=\alpha$ towards $\bar{V}^{\alpha}_t$ solution of the well-posed system of $P$ mean-field equations:
\vspace{-0.2cm}
\begin{equation*}
\dot{\bar{V}}^{\alpha}_t=-\frac {\bar{V}^{\alpha}_t} {\ta}  +I_{\alpha}(t)+\sum_{\beta=1}^P (\bar{J}_{\alpha\beta}+\sigma \dot{W}^{\alpha\beta}_t) \mathbb{E} \big [S(\bar{V}^{\beta}_t)\big]
\end{equation*}

\vspace{-0.3cm}

\noindent where $\dot{W}^{\alpha\beta}_t$ are independent white noise and $\mathbb{E}[\cdot]$ denotes the statistical expectation. Using an implicit form of this equation given by the variation of constant formula, we prove that this system admits Gaussian solutions attracting exponentially fast every solution. A striking property is that the mean and variance of this Gaussian solution satisfy a closed system of ODEs. In details, noting $\mu_{\alpha}$ the mean of the Gaussian solution related to population $\alpha$, $v_{\alpha}$ its variance and $f(\mu,v)$ the expectation of $S(X)$ for $X\sim\mathcal{N}(\mu,v)$ \footnote{$f(\mu,v)=\int_{\R} f(x\sqrt{v}+\mu) Dx$ where $Dx$ is the standard Gaussian density. In particular if $S(x) = \erf(g\,x+\gamma)$, $f(\mu,v)=\erf((g\mu+\gamma)/\sqrt{1+g^2v})$}, we have:
\vspace{-0.2cm}
\begin{eqnarray}
	\label{eq:Mean}\dot{\mu_{\alpha}}=-\frac 1 {\ta} \mu_{\alpha}+ \sum_{\beta=1}^P \bar{J}_{\alpha\beta} f(\mu_{\beta},v_{\beta})+I_{\alpha}(t)\\
	\label{eq:var} \dot{v_{\alpha}}=-\frac 2 {\ta} \,v_{\alpha} + \sigma^2 \sum_{\beta=1}^P f(\mu_{\beta},v_{\beta})^2.
\end{eqnarray}

\vspace{-0.2cm}

\noindent The very complex stochastic mean-field equation thus exactly reduces to a mere system of coupled nonlinear ODEs characterizing the permanent regime. Moreover, the level of disorder appears as a parameter, which allows precise identification of noise-induced transitions using bifurcation theory. 

\begin{figure}[!ht]
	\begin{center}
		\includegraphics[width=.4\textwidth]{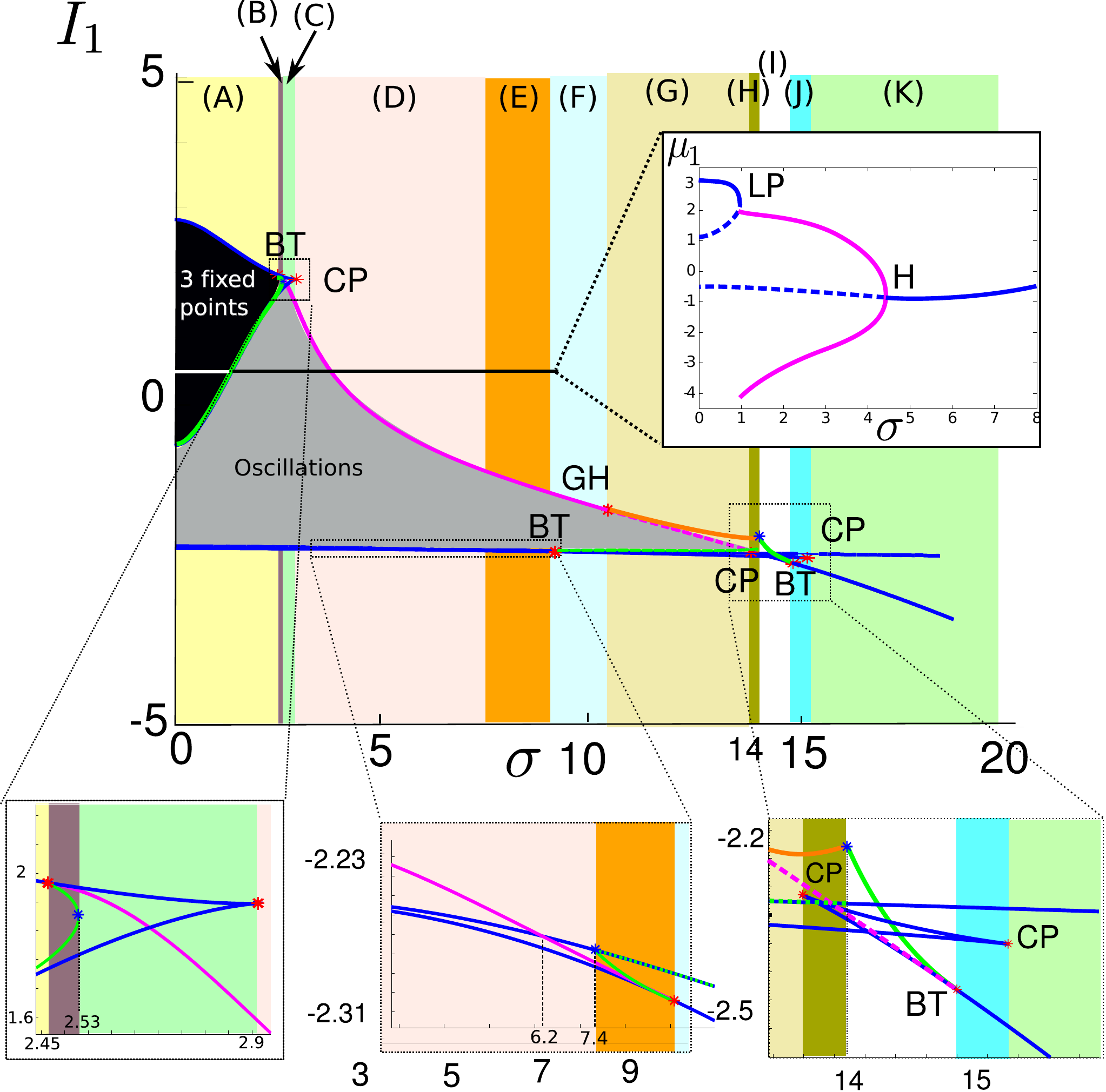}
	\end{center}
	\caption{\small Bifurcation diagram of the moment equations in $\sigma$ and $I_1$. $\bar{J}_{11}=15$, $\bar{J}_{12}=-12$, $\bar{J}_{21}=16$, $\bar{J}_{22}=-5$, $I_2=-3$ following~\cite{wilson-cowan:72}, $\tau_{\alpha}=1$, $S(x)=\erf(x)$. Black region: 3 fixed points, Gray: cycles. Vertical bands: $11$ typical regimes (as $I_1$ is left free) depending on $\sigma$ . Blue: saddle-node, pink: Hopf, plain (resp. dashed) green: saddle (resp saddle-node)-homoclinic, orange: fold of limit cycles (FLC). BT: Bogdanov Takens. CP: Cusp. GH: Bautin. Upper-right inset shows disorder-induced oscillations for $I_1=0$: Plain (dashed) blue: stable (unstable) equilibria, pink: periodic orbits, LP: saddle-node, H: Hopf.}
	\label{fig:Codim2_TwoPops}
\end{figure}

The bifurcation diagram of these mean-field equations in an excitatory/inhibitory case with constant input $I_{\alpha}$ shows a very intricate structure. As a function of $\sigma$ and the external current $I_1$ (Fig.~\ref{fig:Codim2_TwoPops}), it includes local bifurcations (Hopf, fold, Bogdanov-Takens, cusp and Bautin) and global bifurcations (homoclinic and fold of limit cycles) induced by the disorder, defining $11$ regions for $\sigma$ in which the system responds in the same manner to the external input $I_1$ \footnote{The precise analysis of the $10$ heterogeneity-induced bifurcations will appear elsewhere.}. In particular, there exists an interval of $I_1$ for which the disorder induces transitions from a stationary regime to  stable periodic orbits (upper-right inset of Fig.~\ref{fig:Codim2_TwoPops}). When increasing $\sigma$, these limit cycles appear through a saddle-homoclinic bifurcation and disappear through a Hopf bifurcation.

\begin{figure}[htbp]
	\centering
\includegraphics[width=.45\textwidth]{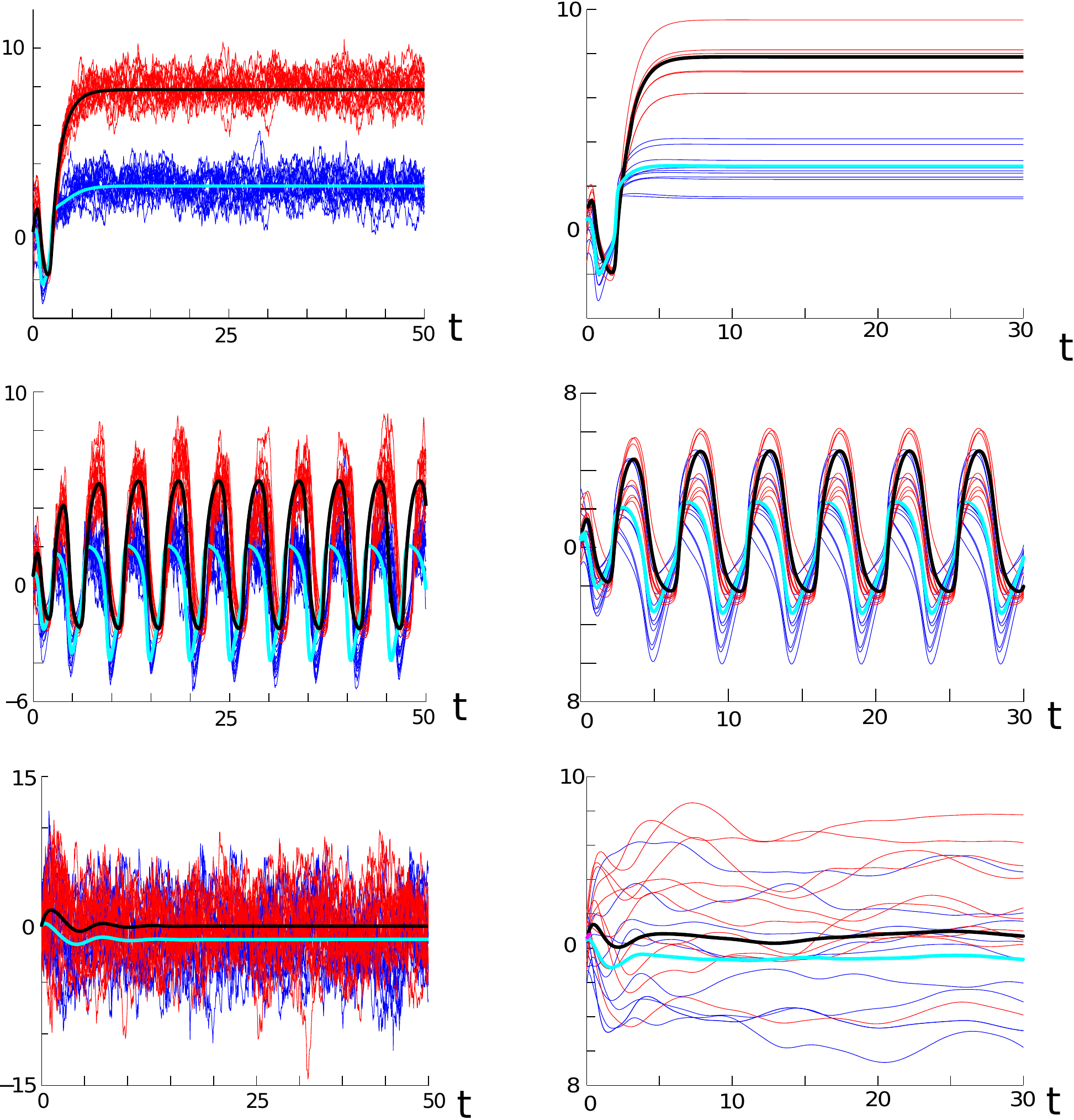}
                \caption{\small Network dynamics in the stochastic noise case (left) and quenched heterogeneity case (right), for $I_1=0$ and increasing values of the disorder parameter $\sigma$ : (left): $0.5$, $1.5$ and $6$, (right): $0.9$, $1.6$ and $3$. $5\,000$ neurons per population, 10 sample trajectories (red: population 1 and blue: population 2) and macroscopic statistics (black: population 1 and cyan: population 2). }
          \label{fig:Behaves}
\end{figure}

This analysis proves that the solution of the stochastic mean-field equations is perfectly periodic in law in a specific interval $\mathcal{I}$ of the disorder parameter, and stationary for smaller or larger heterogeneity. From the microscopic viewpoint, the propagation of chaos property ensures that all neurons in a given population have the same law, and hence present phase-locked oscillations for disorder levels in $\mathcal{I}$ (see Fig.~\ref{fig:Behaves}). 

With this example in mind, we now turn to the quenched heterogeneities case. In the thermodynamic limit, using the formalism developed in the dynamical mean-field theory for spin-glasses~\cite{sompolinsky-zippelius:82,guionnet:97}, the solution of the network equations converge towards the solution of the non-Markovian $P$ implicit well-posed equation:
\vspace{-0.3cm}
\begin{equation}\label{eq:Benarous}
	\dot{\tilde{V}}^{\alpha}_t=-\frac{\tilde{V}^{\alpha}_t}{\tau_{\alpha}}+\sum_{\beta=1}^{P} U_{\alpha \beta}^{\tilde{V}^{\beta}}(t)+I_{\alpha}(t).
\end{equation}

\vspace{-0.3cm}

\noindent In these equations appear the effective interaction processes $(U_{\alpha \beta}^{\tilde{V}^{\beta}}(t))_{\alpha\beta}$ encapsulating the randomness of the synaptic coefficients. These are independent Gaussian processes with mean $\bar{J}_{\alpha\beta} \,\mathbb{E}[S(\tilde{V}^{\beta}_t)]$ and covariance $\sigma^2 \mathbb{E}\Big[S(\tilde{V}^{\beta}_t)S(\tilde{V}^{\beta}_s)\Big]$. Similar to the stochastic synaptic noise case, due to the intrinsic linear dynamics, any solution of~\eqref{eq:Benarous} is exponentially fast attracted towards Gaussian solutions. The mean of Gaussian solutions satisfies exactly the same equation as in the previous case (equation~\eqref{eq:Mean}). Unfortunately in the quenched heterogeneity case, we cannot describe the variance dynamics through a simple ODE. It depends on the covariance of $\tilde{V}^\alpha$ through the implicit integral equation:
\begin{multline*}
C_{\alpha}(t,s)=
e^{-(t+s)/\ta}\Big[e^{2t_0/\ta}C_{\alpha}(t_0,t_0)+\\
\sigma^2\sum_{\beta=1}^P \int_{t_0}^t\int_{t_0}^s e^{(u+v)/\ta}\mathbb{E}\Big[S(\tilde{V}_u^{\beta})S(\tilde{V}_v^{\beta})\Big]dudv\Big]
\end{multline*}
The dependence of $C_{\alpha}(t,s)$ on the whole past of the solution is fundamentally related to the non-Markovian nature of the equations. This more complicated framework does not allow analytical study of the equilibria and their stability. 

However, the fact that the solutions of the mean-field equations, for both stochastic and quenched synaptic heterogenetities, have a mean characterized by the same equation suggests that quenched heterogeneity-induced oscillations may arise. And indeed, it is easy show that the variance of the solution $v_{\alpha}(t)=C_{\alpha}(t,t)$ is an increasing function of the disorder parameter $\sigma$. Artificially considering the variance as a parameter in equation~\eqref{eq:Mean}, routine bifurcation analysis shows that the excitatory/inhibitory system considered undergoes, as $v$ is increased, a saddle-homoclinic bifurcation yielding oscillations that disappear through a Hopf bifurcation~\cite{touboul-hermann-faugeras:11}. Increasing the disorder parameter in the system~\eqref{eq:Benarous} will hence lead the variance to exceed the value related to the saddle-homoclinic bifurcation, yielding a transition from stationary to periodic solutions, that correspond to synchronized oscillations at the microscopic level because of the identical distribution of all neurons in the thermodynamic limit. 
The disorder-induced transition is precisely observed in the numerical simulations of large networks as displayed in Fig.~\ref{fig:Behaves}. Two transitions, first from stationary solutions (weak heterogeneity) to synchronized oscillations and then back to a stationary regime, are observed as heterogeneity increases. 

Beyond the striking statistical similarities, the very different underlying microscopic settings of stochastic and quenched heterogeneity correspond to distinctive qualitative features. In particular, when periodic solutions lose stability, the synaptic noise networks displays a pure stochastic dynamics whereas in the quenched heterogeneity case the network displays a chaotic dynamics, precisely related to the same phenomenon as in~\cite{sompolinsky-crisanti-etal:88} which may be explained through the analysis of the eigenvalues of the random connectivity matrix.

\begin{figure}
	\centering
		\includegraphics[width=.4\textwidth]{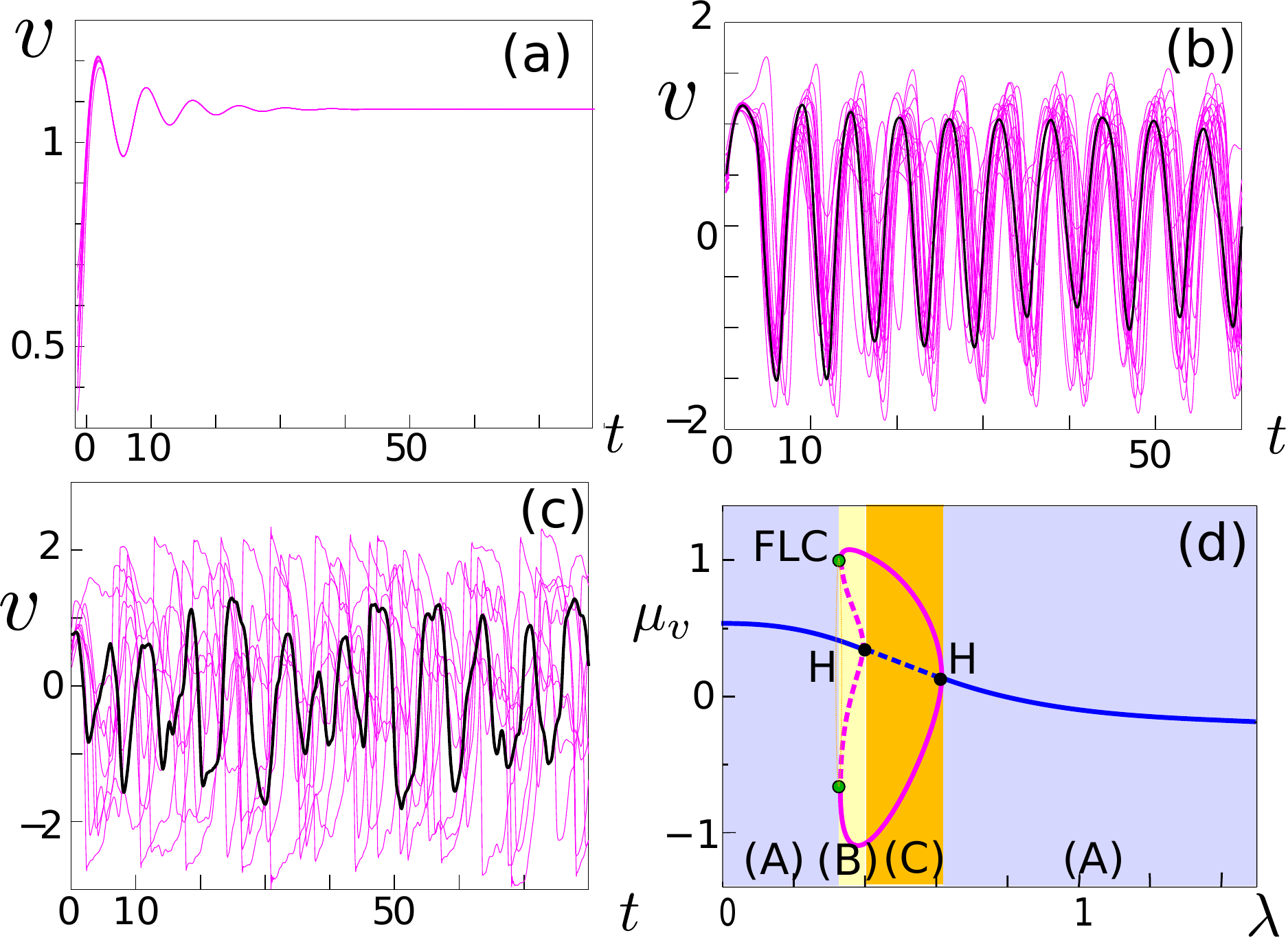}
	\caption{\small FN network, $a=0.4,\, b=2,\, I=0.5, \bar{J}=1.5, \kappa=2$. $20$ trajectories of the voltage chosen among $5\,000$ neurons (pink) (thick black: one sample trajectory). (a) weak heterogeneities ($\sigma=0.5$): ``deterministic'' behavior, (b) intermediate ($\sigma=1$): synchronized oscillations and (c) strong heterogeneities ($\sigma=1$): chaotic behavior. (d) bifurcation diagram of system~\eqref{eq:MomentsFN}: (A): stationary solutions, (B): bistability, (C): oscillations.}
	\label{fig:FHN}
\end{figure}

One clear advantage of firing-rate models are their relative simplicity, allowing theoretical approaches to uncover the phenomenon of disorder-induced transitions. A key ingredient allowing the analytical developments was the linearity of the intrinsic neuronal dynamics, yielding Gaussian solutions to the mean-field equations. Though valid to reproduce a large body of experimental results, this linearity neglects the prominent excitable nature of neurons. We now show that the same transition occurs in a realistic one population network composed of Fitzhugh-Nagumo (FN) neurons interconnected with random electrical synapses. This model describes the dynamics of neuron $i$ through its membrane potential $v^i$ and a recovery variable $w^i$:
\[\begin{cases}
	\dot{v}^i_t=f(v^i_t)-w^i_t+\sum_{j=1}^N J_{ij}(v^j_t-v^i_t)+I\\
	\dot{w}^i_t=a(b\,v^i_t-w^i_t).
\end{cases}\]
In this equation, $a$ represents the time-scale of the recovery variable, $f(v)=v(1-v)(v-\kappa)$ and $I$ is a deterministic input. We choose parameters corresponding to the case where FN neurons present a unique stable equilibrium $(v^*,w^*)$, and consider the synaptic heterogeneity case with Gaussian weights of positive mean $\bar{J}$ and standard deviation $\sigma$.  In that case, a trivial solution of the network equations (whatever $J_{ij}$) is given by $(v^i,w^i)=(v^*,w^*)$ for all $i$. This fixed point is stable for weak heterogeneity (see Fig.~\ref{fig:FHN}). As the disorder parameter $\sigma$ is increased above a critical value, this stationary regime loses stability and periodic synchronized oscillations appear. As the disorder is further increased, these oscillations progressively disappear in favor of a chaotic regime of individual neurons in which the empirical mean becomes stationary. 
Analytical studies in that case are way harder than in the firing-rate case. \reviewThree{Moment expansion and truncation may allow approximating the random system into a deterministic system of ODEs~\cite{zaks}.} Motivated by the previous study, we propose to approximate the solutions by Gaussian processes with averaged voltage $\mu_v$ and recovery variable $\mu_w$ and assuming constant standard deviation $\lambda$ on the voltage. Under this approximation, the variables satisfy the ODE:
\vspace{-0.2cm}
\begin{equation}\label{eq:MomentsFN}
	\begin{cases}
		\dot{\mu}_v=f(\mu_v)+\lambda^2(1+\kappa+3\mu_v)-\mu_w+I\\
		\dot{\mu}_w=a(b\mu_v-\mu_w)
	\end{cases}
\end{equation}

\vspace{-0.2cm}

\noindent and this system shows a transition towards oscillations as $\lambda$ is increased (occurring when $\sigma$ is increased), as plotted in Fig.~\ref{fig:FHN} (d). This transition towards oscillations in the Gaussian approximation allows accounting for the heterogeneity-induced transition to oscillations observed of the original network.  

We hence showed that perfectly periodic phased-locked oscillations arose in the mean-field limit of large scale networks, induced by the presence of heterogeneous connectivity weights. Because of the large amplitude of the oscillations (appearing through homoclinic or after a fold of limit cycle), finite-size networks, converging in $O(1/\sqrt{N})$ towards the mean-field equations, will always display the same kind of perfectly periodic synchronized oscillations as soon as $1/\sqrt{N}$ is small compared to the amplitude of the cycle. \reviewThree{Both nonlinearities and network structure are implicated in this phenomenon: nonlinearity couple the dynamics of the ensemble average to the degree of heterogeneity, and negative feedback loops (inhibitory neural population in the firing-rate model and adaptation variable in the FN model)}. This phenomenon is distinct of more usual noise-induced transitions studied in the field of nonlinear systems with noise (\cite{lindner:04,sagues} and references therein). It distinguishes from stochastic resonance since there is no periodic input, and from coherence resonance since here the mean and standard deviation are perfectly periodic with phase determined by the initial condition. But the main difference with these previous works is the fact that all these phenomena were identified for external noise whereas here the phenomenon evidenced is related to the level of heterogeneities. \reviewThree{In the quenched heterogeneity case, outstanding works addressed the problem of chaotic or order-disorder transitions~\cite{sompolinsky-crisanti-etal:88,toral} or the dynamics of quenched randomly in coupled oscillators~\cite{ullner}, both differing from the transition reported here since we do not deal with coupled oscillators, and addressed a transition to synchronized oscillations.}\\
A sudden apparition of synchronized large-amplitude arbitrarily slow oscillations is strongly evocative of epileptic seizures. These findings provide a theoretical explanation to the recent experimental results that reported persistently increased synaptic variability, without changes in the mean, after febrile seizures in developing rats~\cite{aradi-soltesz:02}. \\
Several intriguing questions remain unanswered. First is the heuristic understanding of the systematic apparition of the oscillations in the presence of heterogeneities. Second is the striking parallel between stochastic networks and randomly connected quenched networks. Besides this particular transition from stationary to periodic solutions, a number of more complex bifurcations were presented here in the synaptic noise case \reviewThree{and a transition to chaotic individual trajectories exhibited in the quenched case and will be examined further and may provide deeper understanding of such heterogeneous networks}. Eventually, the dynamics of such networks with correlated synaptic weights, or with a non-recurrent connectivity graph, are deep questions that are still largely left unexplored.\\
{\small {\bf Acknowledgement:} We acknowledge Paul Bressloff for insightful discussions. GH was partially funded by ERC grant \#227747 NerVi and EU project BrainScaleS.}
\vspace{-0.5cm}
%

\end{document}